# Photometric Studies of Comet C/2009 P1 (Garradd) before the Perihelion

**A. V. Ivanova[a], S. A. Borisenko[a], and M. V. Andreev[a, b, c]**

[a] *Main Astronomical Observatory, National Academy of Sciences of Ukraine, Goloseevo, Kyiv, 03680 Ukraine*

*e-mail: sandra@mao.kiev.ua; IvanovAleksandra@gmail.com*

[b] *International Centre of Astronomical, Medical, and Ecological Research, ul. Akademika Zabolotnoho 27, Kyiv, 03680 Ukraine*

[c] *Terskol Branch of Astronomy Institute, Russian Academy of Sciences, Elbrus region, Karachai-Cherkessian Republic, 361603 Russia*

**Abstract**—The results of the photometric observations of comet C/2009 P1 (Garradd) performed at the 60-cm Zeiss-600 telescope of the Terskol observatory have been analyzed. During the observations, the comet was at the heliocentric and geocentric distances of 1.7 and 2.0 AU, respectively. The CCD images of the comet were obtained in the standard narrowband interference filters suggested by the International research program for comet Hale-Bopp and correspondingly designated the "Hale–Bopp (HB) set." These filters were designed to isolate the BC (λ4450/67 Å), GC (λ5260/56 Å) and RC (λ7128/58 Å) continua and the emission bands of $C_2$ (λ5141/118 Å), CN (λ3870/62 Å), and $C_3$ (λ4062/62 Å). From the photometric data, the dust production rate of the comet and its color index and color excess were determined. The concentration of $C_2$, CN, and $C_3$ molecules and their production rates along the line of sight were estimated. The obtained results show that the physical parameters of the comet are close to the mean characteristics typical of the dynamically new comets.

## INTRODUCTION

Comet C/2009 P1 (Garradd) was discovered with the use of the 0.5-m Schmidt telescope of the Siding Spring Observatory on August 13, 2009 (McNaught and Garradd, 2009), as an object of 17.5*m* in stellar magnitude. In that time, the comet was at the heliocentric distance of 8.7 AU. The orbital parameters of the comet, such as the eccentricity $e = 1.00099925$, the Tisserand parameter $T_j = 0.432$, the inclination $i = 106.2°$, and the semimajor axis $a = 2564.1$ AU (Nakano, 2012), allow us to rate it as a long-period comet originating from the Oort cloud.

The comet passed its perihelion on December 23, 2011, at the heliocentric distance of 1.55 AU and became an object of thorough investigation in different wavelength ranges.

The study of the comet in the infrared range at a heliocentric distance larger than 2 AU revealed traces of such compounds as CO, $CH_3OH$, $CH_4$, $C_2H_6$, and HCN, as well as the hard-to-notice presence (the upper limit of accuracy) of $NH_3$, $C_2H_2$, HDO, and OCS (Paganini et al., 2012). Moreover, from the analysis of individual spatial profiles of the volatile compounds and the continuum, the authors found a significant asymmetry of the material outflow. These results may point to the existence of two separate sources (regions), from which $H_2O$ and CO gases sublimate in different directions. The authors also found a high content of CO, which suggests that comet C/2009 P1 is CO-enriched.

Similar results on the CO abundance in the comet were obtained with the NASA Fund Infrared Telescope (IRTF) at the Mauna Kea Observatory, where the comet was observed before and after the perihelion (McKay et al., 2012).

When comet C/2009 P1 (Garradd) was at the distance of 2.1 AU, observations fulfilled in the infrared range at the Keck II and IRTF telescopes showed the significant water excess in the direction toward the Sun. This can be explained by the sunward water jet containing ice-grain inclusions (Villanueva et al., 2012).

Observations carried out with large telescopes during 2011–2012, when the comet was at the heliocentric distances from 2.40 to 1.57 AU, allowed nine main volatiles to be identified before the perihelion, and six, after the perihelion. The analysis of the data obtained yielded the presence of such compounds as $H_2O$, CO, $CH_4$, $C_2H_2$, $C_2H_6$, HCN, $NH_3$, $H_2CO$, and $CH_3OH$ (Mumma et al., 2012). The authors also found that the comet is CO-enriched, while it is strongly depleted of $C_2H_2$. At the same time, the abundance of $C_2H_6$ and $CH_3OH$ is close to the values measured in the currently observed comets of the Oort cloud.

When comet C/2009 P1 (Garradd) was at the heliocentric distance of 2 AU after passing the perihelion, it was investigated in the infrared range (1.05–4.85 μm) by the *Deep Impact* spacecraft. Three dominant volatile components—$H_2O$, $CO_2$, and CO—were simultaneously identified. The preliminary analysis showed the high abundance of $CO_2$ and CO in the comet relative to that in the other comets (Feaga et al., 2012).

With the *Herschel* space telescope, the D/H content in comet C/2009 P1 was analyzed (Bockelée-Morvan et al., 2012); it turned out to be substantially higher than that for the terrestrial oceans, while the $^{16}O/^{18}O$ content corresponds to the terrestrial estimates.

The spectral observations of the comet carried out in the ultraviolet range with the *Hubble Space Telescope* (HST) in January 2012, also revealed the CO abundance of about 20% relative to that of water (Feldman et al., 2012).

To study the structure and fragmentation inside the cometary nucleus, the morphology of the image of the comet was analyzed. In July 2011, the surface brightness was mapped pixel by pixel with the 0.35-m PARI instrument mounted in the main focus of the HST. The preliminary analysis suggests that the cometary nucleus is a one-piece body of elongated shape. However, the comet shows the signs of probable future fragmentation of the nucleus (Mehnert et al., 2012).

In the present paper, we analyze the photometric data obtained in the observations of comet C/2009 P1 (Garradd) before its perihelion with the use of the HB narrowband cometary filters.

## OBSERVATIONS AND PROCESSING

Comet C/2009 P1 (Garradd) was observed from October 28 to November 14, 2011. During the observations, the comet was at the heliocentric distance $r$ = 1.7 AU and the geocentric distance $\Delta$ = 2.0 AU. The observations of the comet were carried out at the 60-cm Zeiss-600 telescope of the Peak Terskol observatory. The CCD PixelVision Vienna camera with a matrix of 1024 × 1024 pixels was used as a radiation detector. The detector's field of view was 10.7′ × 10.7′, and the image scale was 0.63″ per pixel. The images of the comet were taken with the HB narrowband filters (Farnham et al., 2000) designed to separate the BC, GC, and RC continua and the $C_2$, CN, and $C_3$ emissions.

To increase the signal-to-noise ratio, the binning procedure (2 × 2 pixels into one) was used.

The reduction of the obtained data was performed with the Interactive Data Language (IDL) programs (http://ittvis.com/idl). In the preprocessing, the matrix bias was taken into account, the images were cleared of the traces of cosmic-ray particles, and the flat fields and the dark current were accounted for. The HD 164852 star was observed as a photometric standard (Farnham et al., 2000).

To sum up, the images of the comet in different filters for the further analysis, all of the frames should have been transformed in order to take into account the movement of the comet relative to the background stars. For this, the locations of the comet center and the selected field stars were measured. All of the images were shifted taking accounting of the obtained locations, which resulted in the set of images reduced to the common center with the coordinates corresponding to those of the center of the comet's image chosen in one of the frames. In the obtained images, the photometric center of the comet had the same coordinates, and the centers of the stars were shifted from frame to frame. After that, the sky background was taken into account, and the interframe median filtration was applied to all the transformed images. This procedure allowed the signal-to-noise ratio to be increased and the background stars to be partially removed. The next step in the frame processing was to reduce the images to the same coordinate corresponding to the star center (in the same way, as that for the comet center). The interframe median filtration was again applied to the obtained images, and the resulting images could be used for further analysis and, in particular, for the aperture photometry of the background stars.

The in-depth information on the observations is presented in Table 1 and Fig. 1.

## PHOTOMETRIC INVESTIGATIONS OF THE COMET

The stellar magnitude of the comet was determined by the following formula

$$m_c = -2.5\log\left[\frac{I_c(\lambda)}{I_s(\lambda)}\right] + m_{st} - 2.5\log P(\lambda)\Delta M, \qquad (1)$$

where $m_{st}$ is the magnitude of the standard star, $I_s$ and $I_c$ are the observed fluxes (in relative units) from the star and the comet, respectively, $P$ is the atmospheric transparency coefficient, $\Delta M$ is the difference between the air masses of the star and the comet. The atmospheric transparency coefficient for the Terskol peak was taken from the paper by Kulik et al. (2004).

For the aperture photometry of stars, the diaphragm of 10″ radius was used, which is caused by the fact that the half-width of the Gaussian profile for the background stars was about 4″ on average. The residual background of the sky was estimated with the annular aperture.

The estimates of the stellar magnitude of the comet for the observational period from October 28 to November 14, 2011, are given in Table 1.

**Table 1.** The log of observations of comet C/2009 P1 (Garradd) from October 28 to November 14, 2011

| Observation date, UTC[a] | Exposure, s | Air mass | $r$, AU | $\Delta$, AU | Filters | Magnitude |
|---|---|---|---|---|---|---|
| October, 28.6536 | 300 | 1.49 | 1.730 | 1.986 | RC | 9.14 ± 0.04 |
| October, 28.6653 | 180 | 1.59 | 1.730 | 1.986 | GC | 10.08 ± 0.02 |
| October, 28.6721 | 300 | 1.65 | 1.730 | 1.987 | $C_2$ | 8.43 ± 0.06 |
| October, 28.6766 | 300 | 1.70 | 1.730 | 1.987 | $C_3$ | 9.89 ± 0.08 |
| October, 28.6802 | 300 | 1.74 | 1.730 | 1.987 | BC | 10.77 ± 0.01 |
| October, 28.7104 | 300 | 2.18 | 1.730 | 1.987 | CN | 8.98 ± 0.02 |
| October, 29.6379 | 420 | 1.41 | 1.724 | 1.994 | $C_2$ | 8.30 ± 0.06 |
| October, 29.6436 | 420 | 1.44 | 1.724 | 1.994 | CN | 8.49 ± 0.02 |
| October, 29.6486 | 420 | 1.48 | 1.724 | 1.994 | BC | 10.77 ± 0.01 |
| October, 29.6538 | 420 | 1.52 | 1.724 | 1.994 | RC | 9.12 ± 0.04 |
| October, 29.6590 | 420 | 1.56 | 1.724 | 1.994 | GC | 10.11 ± 0.02 |
| October, 30.6441 | 420 | 1.47 | 1.718 | 2.003 | $C_2$ | 8.32 ± 0.06 |
| October, 30.6508 | 420 | 1.52 | 1.718 | 2.003 | CN | 8.52 ± 0.02 |
| October, 30.6563 | 420 | 1.56 | 1.718 | 2.003 | BC | 10.89 ± 0.01 |
| October, 30.6630 | 420 | 1.62 | 1.718 | 2.003 | RC | 9.14 ± 0.04 |
| October, 30.6689 | 420 | 1.68 | 1.718 | 2.003 | $C_3$ | 10.09 ± 0.07 |
| October, 30.6740 | 420 | 1.74 | 1.718 | 2.003 | GC | 10.22 ± 0.02 |
| October, 31.6369 | 420 | 1.44 | 1.713 | 2.010 | $C_2$ | 8.35 ± 0.06 |
| October, 31.6422 | 420 | 1.48 | 1.713 | 2.010 | $C_3$ | 10.03 ± 0.08 |
| October, 31.6475 | 420 | 1.52 | 1.713 | 2.010 | CN | 8.56 ± 0.02 |
| October, 31.6538 | 420 | 1.57 | 1.713 | 2.010 | BC | 10.86 ± 0.01 |
| October, 31.6591 | 420 | 1.61 | 1.713 | 2.010 | RC | 9.08 ± 0.04 |
| October, 31.6625 | 300 | 1.65 | 1.713 | 2.010 | GC | 10.13 ± 0.02 |
| November, 1.6596 | 420 | 1.65 | 1.707 | 2.018 | $C_2$ | 8.25 ± 0.06 |
| November, 1.6646 | 420 | 1.70 | 1.707 | 2.018 | $C_3$ | 10.03 ± 0.08 |
| November, 1.6704 | 420 | 1.76 | 1.707 | 2.018 | GC | 10.18 ± 0.02 |
| November, 1.6755 | 420 | 1.83 | 1.707 | 2.018 | BC | 10.92 ± 0.01 |
| November, 1.6806 | 420 | 1.89 | 1.707 | 2.018 | RC | 9.13 ± 0.04 |
| November, 1.6993 | 420 | 2.21 | 1.707 | 2.018 | CN | 8.74 ± 0.02 |
| November, 5.6561 | 240 | 1.74 | 1.685 | 2.045 | $C_2$ | 8.29 ± 0.06 |
| November, 5.6626 | 120 | 1.81 | 1.685 | 2.045 | BC | 10.57 ± 0.01 |
| November, 5.6702 | 300 | 1.92 | 1.685 | 2.045 | $C_3$ | 9.95 ± 0.08 |
| November, 5.6775 | 300 | 2.03 | 1.685 | 2.045 | RC | 9.18 ± 0.04 |
| November, 14.6497 | 300 | 1.99 | 1.641 | 2.091 | $C_2$ | 8.24 ± 0.06 |
| November, 14.6626 | 300 | 2.22 | 1.641 | 2.091 | BC | 10.99 ± 0.01 |
| November, 14.6597 | 300 | 2.16 | 1.641 | 2.091 | GC | 10.20 ± 0.02 |
| November, 14.6784 | 120 | 2.60 | 1.641 | 2.091 | RC | 9.20 ± 0.04 |

[a] UTC—Coordinated Universal Time.

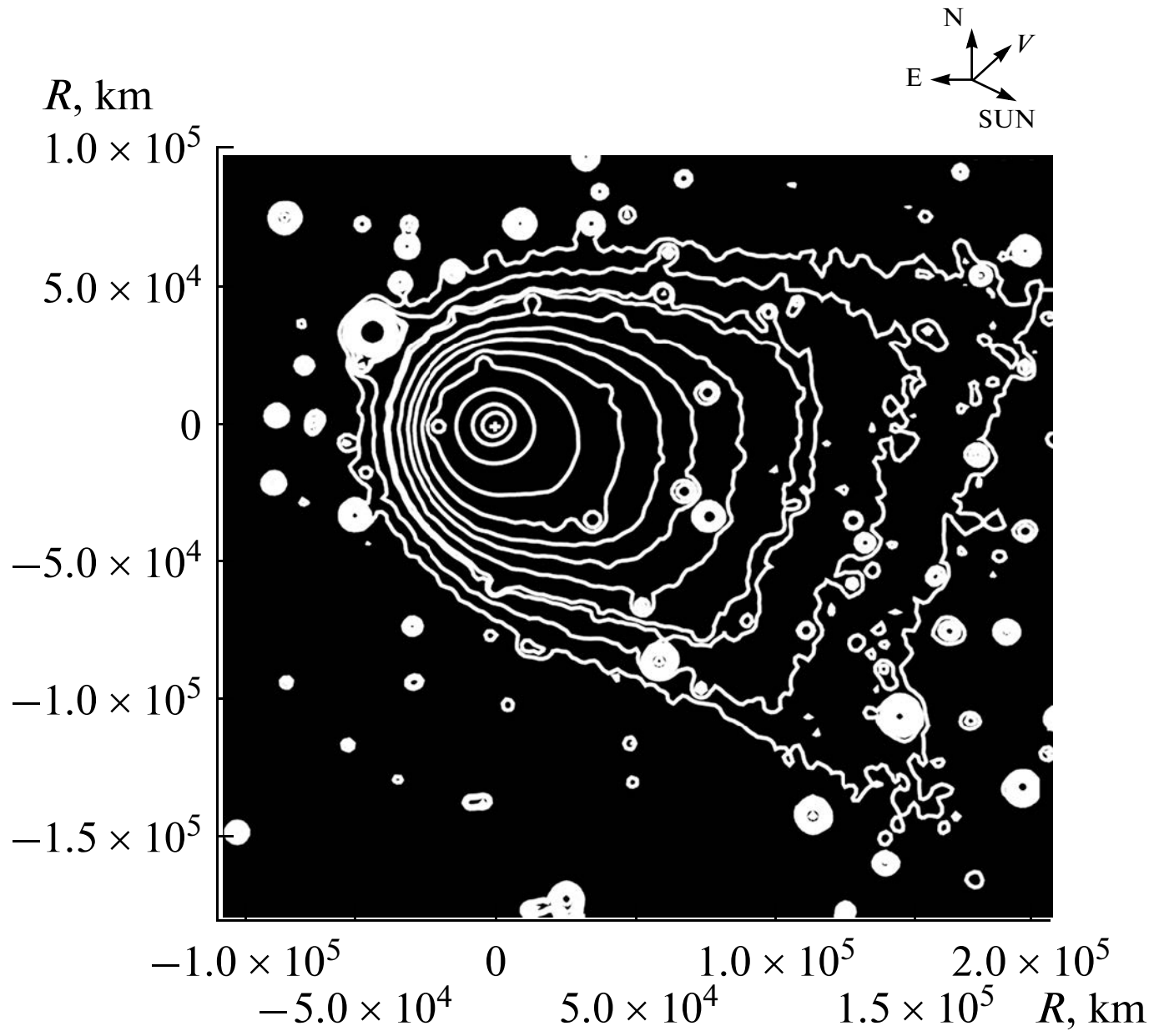

**Fig. 1.** There are the isophots of comet C/2009 P1 (Garradd) for the images obtained in the RC filter on November 14, 2011. They are presented in the distance scale relative to the comet center *R* [km]. The directions to the north, the east, and the Sun, as well as the moving direction of the comet (*V*) are shown.

*The Production Rate and Reflectance Gradient of the Dust and the Color Excess of the Comet*

The analysis of the flux of solar light scattered by the comet allows the following characteristics of dust, such as the dust production rate, the color, the spectral gradient of the reflectance, and the gas-to-dust ratio to be estimated. To determine the column density of dust from the continuum measurements is a rather complicated problem, since the brightness of the comet in continuum depends on the size distribution of particles, the phase angle of the comet, the albedo, etc. Because of this, to estimate the dust production rate in a comet from the flux of its radiation in continuum, the quantity $Af\rho$ is used. This quantity was introduced by A'Hearn et al. (1984), and it is determined by the ratio of the effective cross-section of all particles falling within the receiver's field of view to the projection of the field of view onto the celestial sphere. To estimate this quantity, the following expression is often used (Mazzota Epifani et al., 2010):

$$Af\rho = \left(4r^2\Delta^2 10^{0.4(m_s - m_{cf})}\right)/\rho, \qquad (2)$$

where the quantity $Af\rho$ is expressed in centimeters; $\Delta$ is the geocentric distance also expressed in centimeters; $r$ is the heliocentric distance expressed in astronomical units; $\rho$ is the radius of the aperture used for integrating the signal from the comet projected onto the celestial sphere and expressed in centimeters, i.e., corresponding to the cometocentric radii; $m_s$ and $m_{cf}$ are the stellar magnitudes of the Sun and the comet, respectively. The magnitude of the Sun in the BC, GC, and RC filters was taken from the paper by Farnham et al. (2000).

Formula (2) is often used for estimating the dust production rate in spite of the fact that the condition of isotropic scattering on dust particles in the cometary atmospheres (assumed as a basis for the expression) is evidently not satisfied. To account for possible anisotropic outflows from the cometary nucleus in the calculations of the dust production rate, the stellar magnitude is estimated within a small aperture covering the near-nucleus area, where the condition of the homogeneous outflow of the material from the cometary nucleus is assumed to be satisfied. In Fig. 2, the logarithmic dependence of the parameter $Af\rho$ on the heliocentric distance of the comet is presented.

To characterize quantitatively the scattering by dust at different wavelengths, the spectral gradient of the reflectance is used (Jewitt and Meech, 1986). The spectral gradient is expressed in percents per 1000 Å, which allows the difference between the filters that are used for isolating the continuum in the comets to be taken into account

$$S'(\lambda_1, \lambda_2) = \frac{2000}{\Delta\lambda}(I_{red} - I_{blue})/(I_{red} + I_{blue}), \qquad (3)$$

where $\Delta\lambda$ is the difference between the effective wavelengths of the red and blue filters expressed in angstroms and $I_{red}$ and $I_{blue}$ are the radiation fluxes of the comet in the red and blue spectral ranges, respectively. With the use of the above-described technique, we estimated the spectral gradient of the reflectance for the whole period of observations of the comet.

The color excess of comet C/2009 (Garradd) was also estimated; it is characterizes by the quantity

$$CE = C - [m_{\mathrm{Sun}}(\lambda_1) - m_{\mathrm{Sun}}(\lambda_2)], \tag{4}$$

where $C = m_c(\lambda_1) - m_c(\lambda_2)$ is the observed color index of the comet and $m_{\mathrm{Sun}}(\lambda_1) - m_{\mathrm{Sun}}(\lambda_2)$ is the color index of the Sun. The results are presented in Table 2.

### *The Column Density and Production Rate of Molecules*

With the use of narrowband filters for isolating the emission bands, we estimated the column density of molecules and their production rate.

When estimating the total number of molecules in the column of radius ρ along the line of sight, we assume that the cometary atmosphere is optically thin and, consequently, the fluxes are in direct proportion to the number of emitters, which allows us to use the following formula:

$$N = L/g,$$

where $L = 4\pi\Delta^2 F_{\mathrm{cf}}$ is the observed emittance of the comet in the emission band and $F_{\mathrm{cf}}$ is the emission flux of the comet in this band, Δ is the geocentric distance expressed also in astronomic units, and *g* is the fluorescence efficiency for the specified molecule at the heliocentric distance of 1 AU.

In our calculations, we assumed $g = 4.4 \times 10^{-13}$, $1.0 \times 10^{-12}$, and $3.86 \times 10^{-13}$ erg/s mol for the molecules $C_2$ (Newburn and Spinrad, 1989), $C_3$ (A'Hearn, 1982), and CN (obtained with the method described by Tatum (1984)), respectively.

To estimate the production rate, the model by Haser (1957) is used,

$$N(\rho) = \frac{Q}{2\pi V\rho}\frac{\gamma_{\mathrm{d}}}{\gamma_{\mathrm{p}} - \gamma_{\mathrm{d}}}\left[\int\limits_0^{\rho/\gamma_{\mathrm{p}}} K_0(y)\mathrm{d}y - \int\limits_0^{\rho/\gamma_{\mathrm{d}}} K_0(y)\mathrm{d}y\right],$$

where *V* is the escape velocity of molecules, $\gamma_{\mathrm{p}}$ and $\gamma_{\mathrm{d}}$ are the characteristic path lengths of the parent and daughter molecules, respectively, and $K_0$ is the first-order modified Bessel function of the second kind. When calculating the production rate of molecules *Q*, the following characteristic path lengths of the parent and daughter molecules were assumed: $\gamma_{\mathrm{p}} = 2.5 \times 10^4 r^2$ and $\gamma_{\mathrm{d}} = 1.2 \times 10^5 r^2$ for $C_2$ (Cochran, 1985), $\gamma_{\mathrm{p}} = 3.1 \times 10^4 r^2$ and $\gamma_{\mathrm{d}} = 1.4 \times 10^5 r^2$ for $C_3$ (Festou et al., 1990), and $\gamma_{\mathrm{p}} = 1.7 \times 10^4 r^2$ and $\gamma_{\mathrm{d}} = 3.0 \times 10^5 r^2$ for CN, respectively.

The escape velocity of molecules from the nucleus was determined by the formula (Delsemme, 1982)

$$V = 0.58 r^{-0.5}.$$

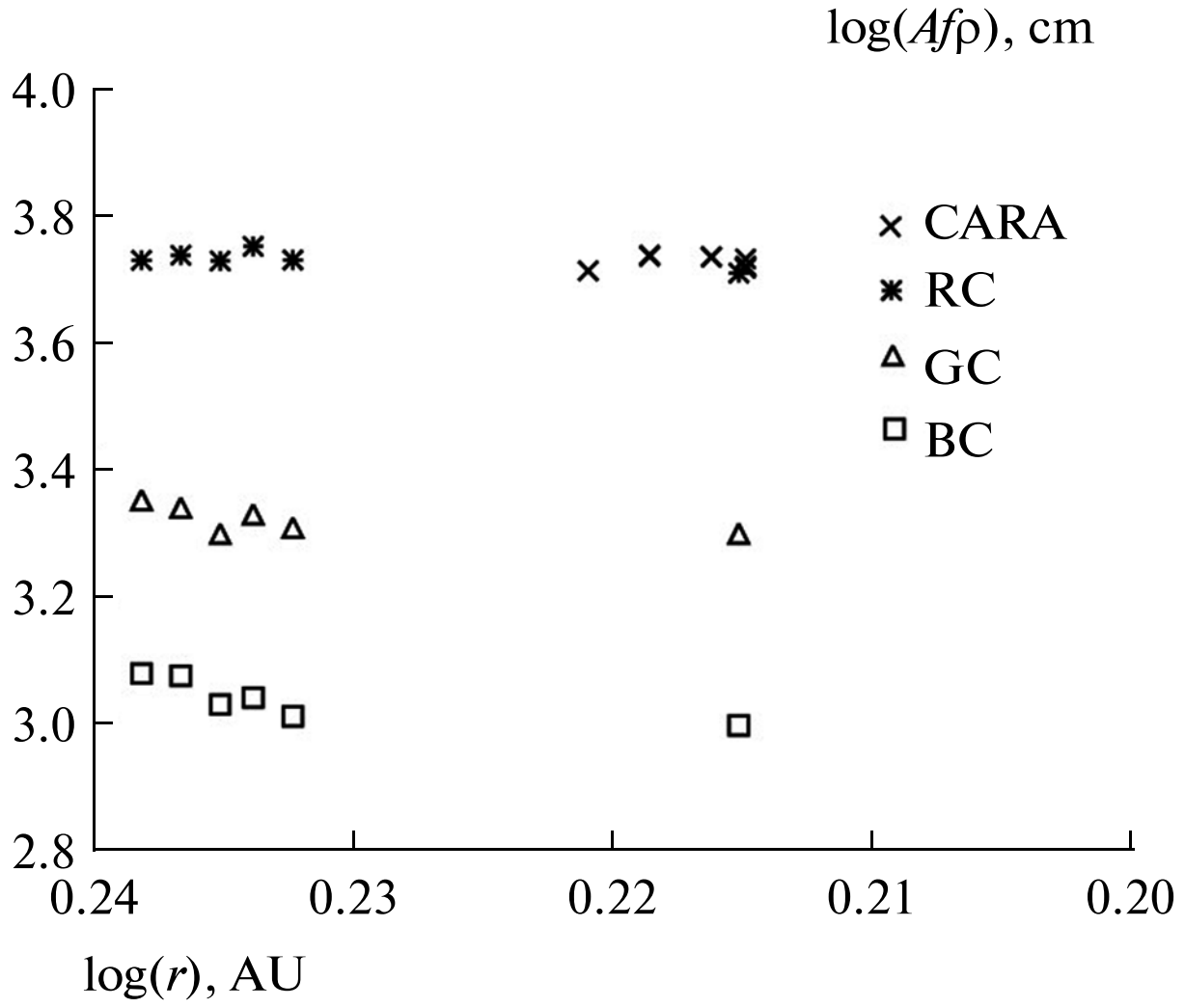

**Fig. 2.** Variations of the parameter *Af*ρ in different filters versus the heliocentric distance of comet C/2009 P1 (Garradd) (the scale is logarithmic). The data for the comet in the R filter taken from the CARA database (Cometary Archive of the *Af*ρ parameter, http://cara.uai.it/) are also presented.

The estimates of the production rate of molecules, their column density, and the dust production rate of comet C/2009 P1 (Garradd) are listed in Table 3.

To characterize quantitatively the gas-to-dust ratio in comets, the ratio of the flux measured in the $C_2$ band to that in the continuum $WC_2 = F_{C2}/F_{BC}$ is used (Krishna Swamy, 2010). For the comets with a strong continuum, the value of the equivalent width of the $C_2$ band is in the range below 500 Å (Kiselev, 2003). We calculated the equivalent widths for comet C/2009 P1 (Garradd), and their values are in Table 3.

## DISCUSSION AND CONCLUSIONS

In the paper, the results of photometric observations of comet C/2009 P1 (Garradd) performed from October 28 to November 14, 2011 are presented.

The analysis of the photometric data showed that the continuum of the comet is redder than the solar one and the color excess ranges, on average, from 0.65 to 1.8 of a stellar magnitude for different cometary filters. For the whole observational period, the value of the spectral gradient of the reflectance varied within the interval from 0 to 20% per 1000 Å, which is characteristic of the most comets (Jewitt and Meech, 1986).

The values of *Af*ρ, which is a measure of the dust production in comet C/2009 P1 (Garradd), varied from 900 to 5500 cm in the continuum filters. From the estimates of the color and the dust production rate, comet C/2009 P1 (Garradd) is close to the dust comets.

**Table 2.** The color index and color excess of comet C/2009 P1 (Garradd)

| Observation date, UTC[a] | $r$, AU | Filters | Color index of the comet, C | Color excess of the comet, CE | S', %/1000 Å |
|---|---|---|---|---|---|
| October, 28.6802 | 1.730 | BC-GC | 0.69 | 0.18 | 2.07 |
| | | GC-RC | 0.94 | 0.17 | 0.84 |
| | | BC-RC | 1.63 | 0.35 | 1.21 |
| October, 29.6486 | 1.724 | BC-GC | 0.66 | 0.15 | 1.74 |
| | | GC-RC | 0.99 | 0.22 | 1.08 |
| | | BC-RC | 1.65 | 0.37 | 1.27 |
| October, 30.6740 | 1.718 | BC-GC | 0.67 | 0.16 | 1.85 |
| | | GC-RC | 1.08 | 0.31 | 1.52 |
| | | BC-RC | 1.75 | 0.47 | 1.60 |
| October, 31.6538 | 1.713 | BC-GC | 0.73 | 0.22 | 2.52 |
| | | GC-RC | 1.05 | 0.28 | 1.38 |
| | | BC-RC | 1.78 | 0.50 | 1.70 |
| November, 1.6755 | 1.707 | BC-GC | 0.74 | 0.23 | 2.64 |
| | | GC-RC | 1.05 | 0.28 | 1.38 |
| | | BC-RC | 1.79 | 0.51 | 1.73 |
| November, 14.6626 | 1.641 | BC-GC | 0.79 | 0.28 | 3.19 |
| | | GC-RC | 1.00 | 0.23 | 1.13 |
| | | BC-RC | 1.79 | 0.51 | 1.73 |

[a] UTC—Coordinated Universal Time.

**Table 3.** Molecular and dust production in comet C/2009 P1 (Garradd) from October 28 to November 14, 2011

| Observation date, UTC | $r$, AU | Δ, AU | $Q$, mol/s | | | $N$, mol/cm$^2$ | | | $Af\rho$, cm | | | $WC_2 = F_{C2}/F_{BC}$, Å |
|---|---|---|---|---|---|---|---|---|---|---|---|---|
| | | | [a]CN | [b]$C_3$ | [a]$C_2$ | [a]CN | [b]$C_3$ | [a]$C_2$ | [c]BC | [c]GC | [c]RC | |
| October, 28.6536 | 1.730 | 1.986 | $5.24 \times 10^{26}$ | $1.01 \times 10^{26}$ | $3.07 \times 10^{26}$ | $1.88 \times 10^{29}$ | $3.00 \times 10^{28}$ | $1.21 \times 10^{29}$ | 1205.3 | 2254.7 | 5359.2 | 5.36 |
| 29.6379 | 1.724 | 1.994 | $8.59 \times 10^{26}$ | – | $3.45 \times 10^{26}$ | $2.97 \times 10^{29}$ | – | $1.37 \times 10^{29}$ | 1194.4 | 2193.5 | 5459.4 | 6.09 |
| 30.6441 | 1.718 | 2.003 | $8.33 \times 10^{26}$ | $9.77 \times 10^{25}$ | $3.34 \times 10^{26}$ | $2.93 \times 10^{29}$ | $2.52 \times 10^{28}$ | $1.39 \times 10^{29}$ | 1076.6 | 1995.6 | 5346.5 | 6.52 |
| 31.6369 | 1.713 | 2.010 | $7.88 \times 10^{26}$ | $9.83 \times 10^{25}$ | $3.22 \times 10^{26}$ | $2.84 \times 10^{29}$ | $2.67 \times 10^{28}$ | $1.33 \times 10^{29}$ | 1104.2 | 2143.2 | 5637.1 | 6.28 |
| November, 1.6596 | 1.707 | 2.018 | $6.46 \times 10^{26}$ | $9.52 \times 10^{25}$ | $3.49 \times 10^{26}$ | $2.38 \times 10^{29}$ | $2.67 \times 10^{28}$ | $1.44 \times 10^{29}$ | 1032.1 | 2040.5 | 5367.0 | 7.35 |
| 14.6497 | 1.641 | 2.091 | – | – | $3.11 \times 10^{26}$ | – | – | $1.45 \times 10^{29}$ | 998.6 | 1998.3 | 5118.6 | 7.98 |

[a] The values were calculated with the aperture radius of 75.2″.
[b] The values were calculated with the aperture radius of 62.7″.
[c] The values were calculated with the aperture radius of 37.6″.

The molecular production rate in comet C/2009 P1 (Garradd) calculated with Haser's model averages $3.3 \times 10^{26}$ mol/cm$^2$ for $C_2$ and amounts to $7.3 \times 10^{26}$ and $9.7 \times 10^{25}$ mol/cm$^2$ for CN and $C_3$, respectively. The values obtained for the molecular production rate are typical of the long-period comets and dynamically new comets (Langland-Shula and Smith, 2011). According to the estimates, for all the observational dates, the ratio $\log(Af\rho(BC)/Q(C_2))$ is about −23.4 on average, which is within the range typical of comets, from −22.11 to −24.83 (Rosenbush, 2004). So, from the overall analysis of the photometric observations, we may conclude that the results allow us to classify comet C/2009 P1 (Garradd) as a typical dust comet.

## ACKNOWLEDGMENTS

We would like to thank N.N. Kiselev for help with the manuscript preparation and useful comments and to the participants of the CARA project (http://cara.uai.it/) A. Milani, G. Sostero, E. Bryssinck, and R. Ligustri for providing the data on the parameter $Af\rho$ for comet C/2009 P1 (Garradd).